\documentclass[12pt,a4paper]{article}%
\usepackage{jheppub}
\pdfoutput=1
\usepackage{amsmath,amssymb,amsfonts,wasysym}
\usepackage{color}
\usepackage{float}
\usepackage[bf,footnotesize]{caption2}
\usepackage[]{hyperref}%
\def\l{\label}
\def\La{\mathcal{L}}

\def\({\left(}
\def\){\right)}
\def\f{\frac}
\def\be{\begin{equation}}
\def\ee{\end{equation}}
\def\s{\sigma}
\def\g{\gamma}
\def\G{\Gamma}

\title{Narrow resonances at the early LHC}
\author{Riccardo Torre}
\affiliation{CERN, Physics Department, Theory Division, CH-1211 Geneva 23, Switzerland}
\affiliation{Dipartimento di Fisica and INFN, Universit\`a di Pisa, Largo Fibonacci 3, I-56127 Pisa, Italy}
\affiliation{Institut f\"ur Theoretische Physik, Universit\"at Z\"urich, Winterthurerstrasse 190, CH-8057 Z\"urich, Switzerland}
\emailAdd{Riccardo.Torre@cern.ch}

\proceeding{}

\proceeding{\vspace{-1.65cm}\flushright{\small CERN-PH-TH/2011-048\\
IFUP-TH/2011-6\\
LPN11-12\\}\vspace{5mm}
\flushleft{\footnotesize Based on a talk given at the workshop\\
10th Hellenic School and Workshops on Elementary Particle Physics and Gravity\\
Corfu, Greece, 29 August - 19 September 2010}}

\abstract{We discuss the possibility to observe new resonances at the early LHC using a phenomenological approach. 
  We show predictions for the production of a scalar in the $gg$-channel and an heavy quark in the $qg$-channel and we compare them with the recent results of the ATLAS and CMS experiments in the di-$jet$ channel, setting upper limits on the relevant couplings. We finally discuss the importance of the final states containing at least one photon making sensitivity predictions for the corresponding couplings.}

\begin{document}
%
\maketitle


\section{Introduction}
The first run of the LHC aims to collect few fb$^{-1}$ of data in the $pp$ collisions at a center of mass energy $\sqrt{s}=7$ TeV. This is the first possibility to explore the Fermi scale beyond the reach of the Tevatron ($\sqrt{s}=1.96$ TeV). It is therefore very important to understand what we can expect during and at the end of this run. It was recently pointed out that with about $5$ fb$^{-1}$ of integrated luminosity, the combination of the data collected from both the ATLAS and CMS experiments can suffice to exclude, or even to discover, the Standard Model (SM) Higgs boson down to the LEP limit of $114$ GeV \cite{CMS:2010Hig,ATLAS:2011Hig}. Even if the discovery of the SM Higgs boson is certainly the most important task of the early LHC, we think that an open attitude in the expectation for possible signals of new physics is necessary at this stage. Model building prejudices normally play an important role in the determination of the experimental strategies. However, we think that these prejudices should be avoided as much as possible in order to develop model independent search strategies and to maximize the sensitivity to a large range of new physics models. For these reasons, we want to discuss the possible signals of new physics that we can expect at the early LHC in a model independent way. We base our considerations on simple phenomenological Lagrangians fulfilling reasonable consistency conditions without aiming to explain the underlying theory from which the states under discussion can arise.\\
\indent The single production of a relatively narrow resonance is the most obvious candidate for a large rate of new physics events. The case of the $q\bar{q}$ vector resonances, either neutral or charged, has been widely studied (see, e.g., Refs.~\cite{Langacker:2008yv, Salvioni:2010p010,Grojean:2011vu}). However, for parton luminosity reasons, we find that at the LHC, in competition with the Tevatron, the $q\bar{q}$-channel is definitely less favorable with respect to the $gg$-, $qg$- and $qq$-channels. Considering for simplicity the lower spin and SM representations there are only few states that can be produced in the channels mentioned above \cite{Barbieri:2011p2759}:
\begin{itemize}
\item {\bf $gg$-channel}: a spinless totally neutral scalar $S$;
\item {\bf $qg$-channel}: a $J=1/2$ color-triplet "heavy quark", either "U" or "D";
\item {\bf $qq$-channel}: a spinless color-triplet or color-sextet $\phi$, with various possible charges.
\end{itemize}

\indent In our phenomenological study we always assume that only a single new particle is available at a time, which therefore can only decay into SM particles. Moreover, we concentrate our attention on the first two cases, since the scalar triplet or sextet can only decay into pair of jets \cite{Bauer:2010p280}, consistenly with known constraints, whereas we find relatively more promising the final states containing at least one photon. The resonaces in the $qq$-channel also suffer of problems with flavor physics \cite{Barbieri:2011p2759}. For an estensive analysis of all the possible colored resonances at the early LHC see Ref.~\cite{Han:2010p2696}.
 
\section{Neutral scalar singlet $S$}
\begin{figure}[t]
\centering
\includegraphics[scale=0.24]{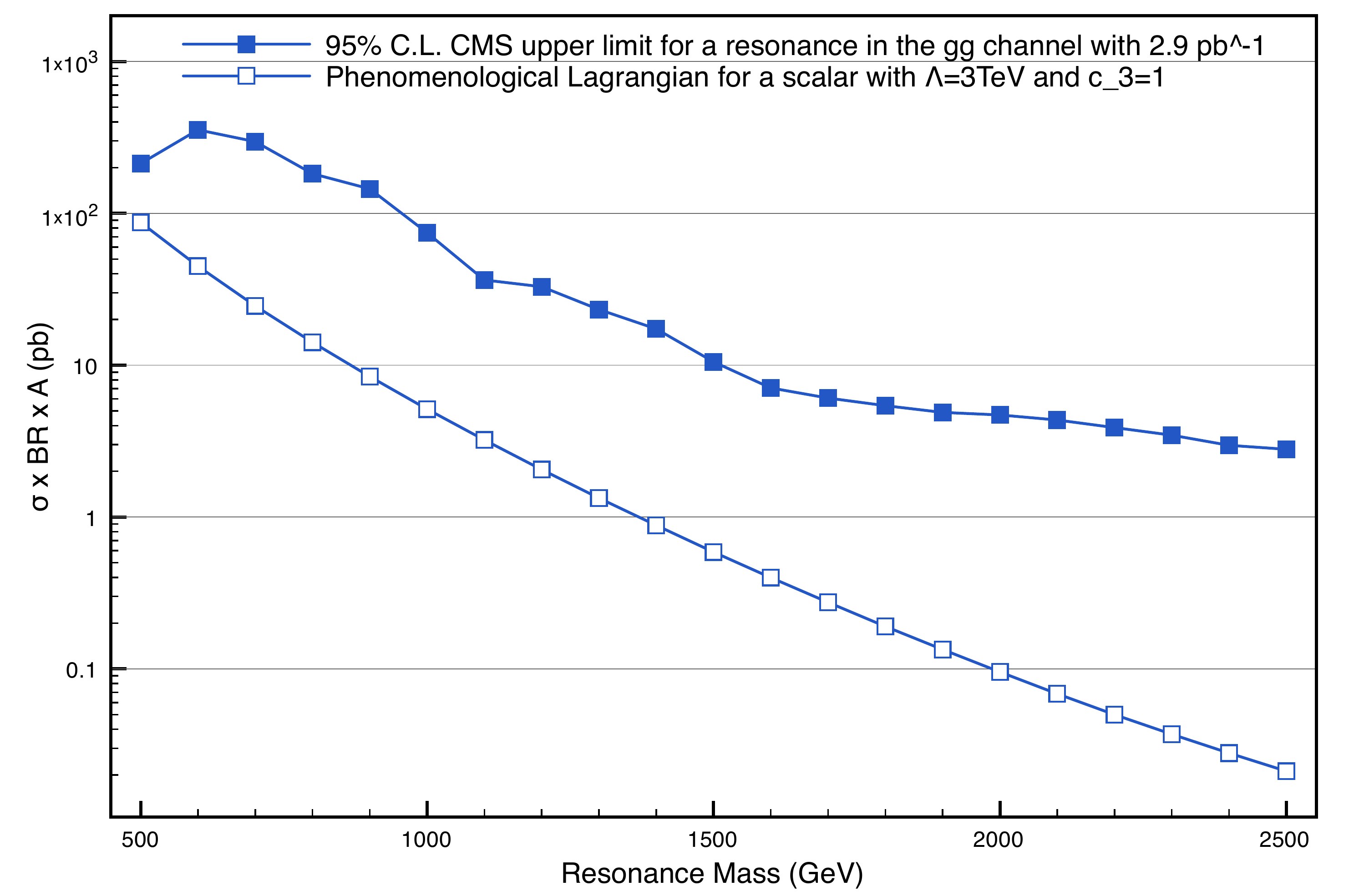}\hspace{3mm}
\includegraphics[scale=0.24]{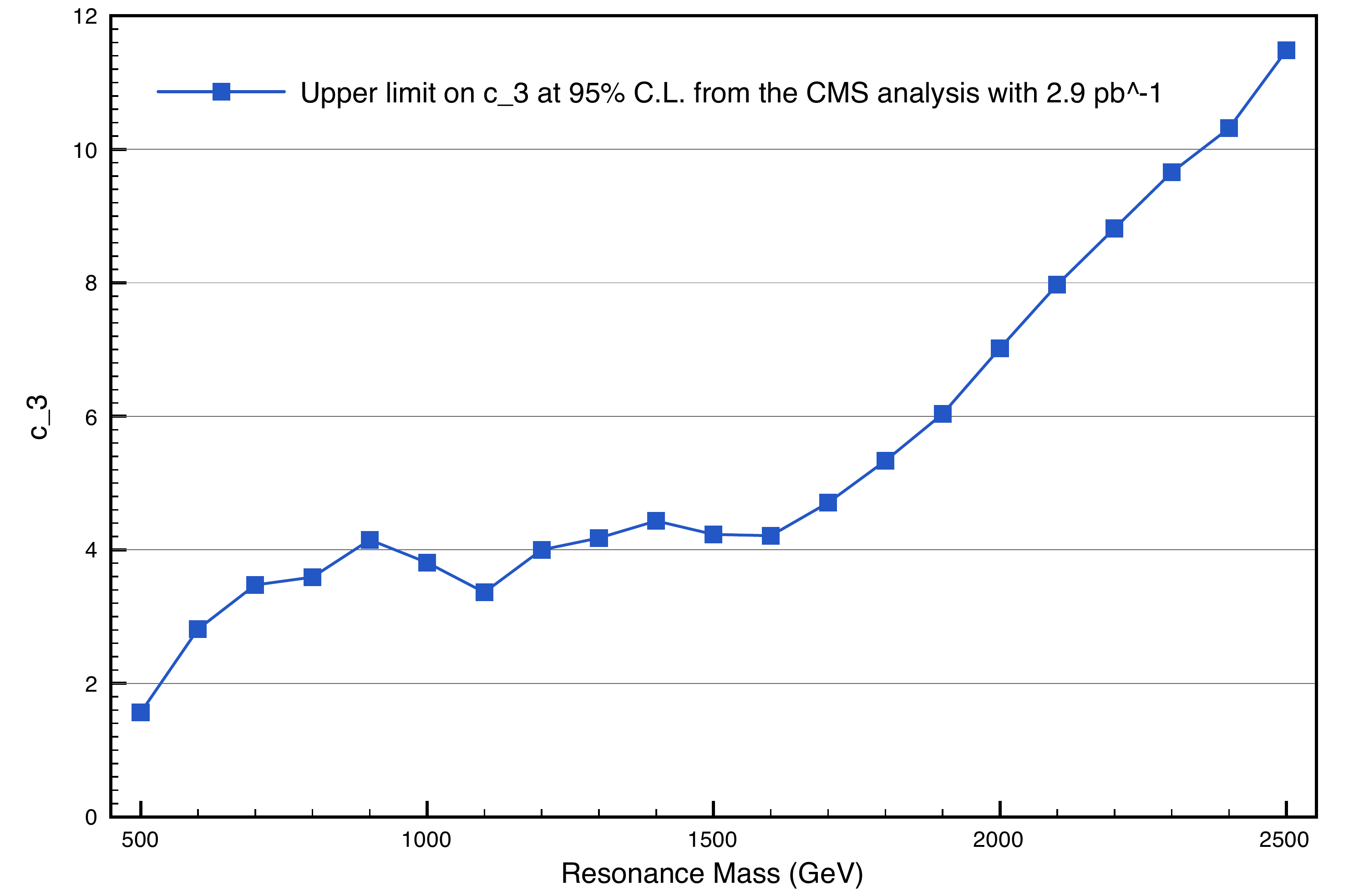}
\caption{Left panel: experimental upper limit at $95\%$ C.L. on the total cross section for a $gg$ resonance measured by the CMS experiment and total cross section predicted with the Lagrangian \eqref{eq1} for the scalar $S$, as functions of the resonance mass; Right panel: upper bound at $95\%$ C.L. on the coupling $c_{3}$ as a function of the resonance mass for $\Lambda=3$ TeV, assuming $BR\(S\to gg\)\approx 1$.}
\label{Scalar_3}
\end{figure}

A reasonable effective Lagrangian describing the interactions of a neutral scalar singlet $S$ of mass $M_{S}$ with the SM particles is
\be\l{eq1}
\La_{S}=c_{3}\f{g_{S}^{2}}{\Lambda}G_{\mu\nu}^{a}G^{\mu\nu\,a}S+c_{2}\f{g^{2}}{\Lambda}W_{\mu\nu}^{i}W^{\mu\nu\,i}S+c_{1}\f{g^{\prime 2}}{\Lambda}B_{\mu\nu}B^{\mu\nu}S+\sum_{f}c_{f}\f{m_{f}}{\Lambda}\bar{f}fS\,,
\ee
where $\Lambda$ is an energy scale, $c_{i}$, $i=1,2,3,f$ are dimensionless constants and $f$ is any SM fermion, of mass $m_{f}$. \\
%
\indent The Branching Ratios (BR) of $S$ are independent on the scale $\Lambda$. 
The total width of $S$ depends on the parameters $M_{S}$, $\Lambda$ and $c_{i}$ and scales as $1/\Lambda^{2}$. We take as reference values $\Lambda= 3$ TeV, reminiscent of a new strong interaction possibly responsible for ElectroWeak Symmetry Breaking (EWSB) at $\Lambda \approx 4\pi v$ and $c_{i}=1$. If $S$ were a composite particle generated by a new strong dynamics, the Naive Dimensional Analysis (NDA) \cite{Georgi:1984p3322} would suggest $c_{i}=1/4\pi$. However, larger values could arise from large $N$ and/or from more drastic assumptions about the nature of the gauge bosons. For the reference values of the parameters we are considering, $\Gamma_{S}$ ranges from about $10$ GeV at $M_{S}=500$ GeV to about $250$ GeV for $M_{S}=1.5$ TeV.\\
\indent We have made a preliminary study of the sensitivity to the search for $S$ in the di-$jet$ and $\gamma\gamma$ final states (see Ref.~\cite{Barbieri:2011p2759} for more details).
We have found that, while the large systematic uncertainties and the low signal over background ratio seem to disfavor an emerging signal in the di-$jet$ channel, at least with the first run integrated luminosity ($1-5$ fb$^{-1}$), a discovery in the $\gamma\gamma$-channel looks possible with a modest integrated luminosity of about $10-100$ pb$^{-1}$.
All these considerations can be trivially extended to arbitrary values of $\Lambda$ and $c_{i}$. In particular, assuming the dominance of the $S\to gg$ channel and the validity of the Narrow Width Approximation\footnote{The NWA allows one to write the total cross section as the product of the production rate times the BR: \mbox{$\s\(pp\to X\to 2x\)\approx \s\(pp\to X\)\times BR\(X\to 2x\)$}.} (NWA), an absence of signal for a given integrated luminosity is easily translated in a lower bound on $\Lambda/c_{3}$ for the di-$jet$ channel and $\Lambda/\(c_{1}+c_{2}\)$ for the $\gamma\gamma$ channel.\\
\indent We have used the recent search for a narrow resonance in the di-$jet$ channel performed by the CMS experiment\footnote{We could not use in this case the corresponding search of ATLAS \cite{ATLASCollaboration:2010p1746} since it was made assuming the $qg$ final state that gives rise to a different quantity of QCD radiation.} \cite{CMSCollaboration:2010p2595} to set an upper bound on the coupling $c_{3}$ for fixed $\Lambda=3$ TeV as a function of $M_{S}$. The upper bound on the total cross section compared with the prediction for our reference values of the parameters and the corresponding bound on the coupling $c_{3}$ are depicted in Fig.~\ref{Scalar_3}.%
\begin{figure}[t]
\centering
\includegraphics[scale=0.26]{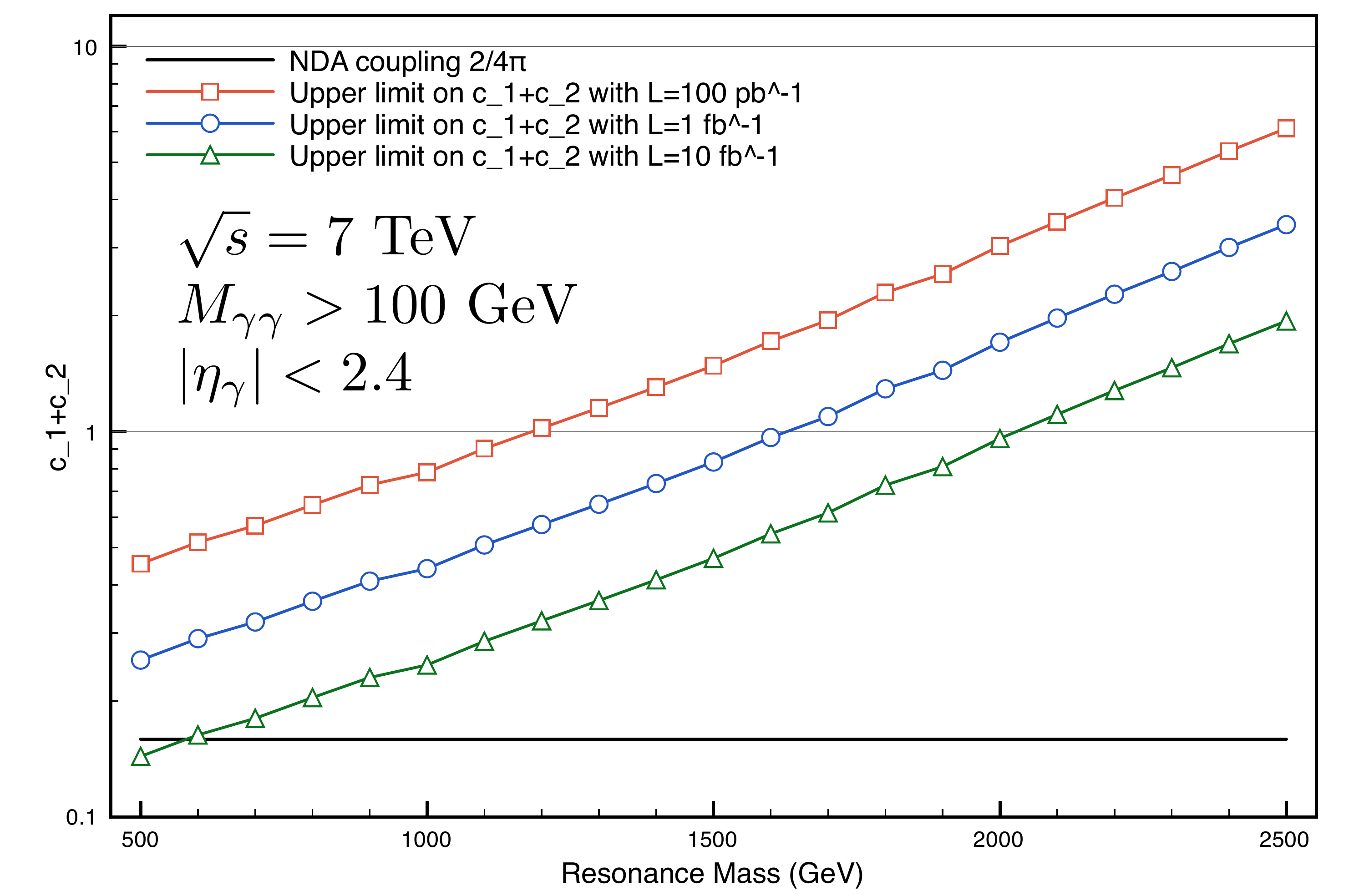}
\caption{Upper limits on $c_{1}+c_{2}$ as functions of the scalar mass $M_{S}$ for $\Lambda=3$ TeV assuming $BR\(S\to \g\g\)\ll BR\(S\to gg\)$.}
\label{Scalar_4}
\end{figure}%
In Fig.~\ref{Scalar_4} we also show a preliminary estimation of the senstivity of the early LHC to the couplings combination $c_{1}+c_{2}$ relevant for the $\gamma\gamma$ final state. From this figure we see that an integrated luminosity of $1$ to $10$ fb$^{-1}$ is required to approach the NDA limit $1/4\pi$ for the couplings $c_{1}$ and $c_{2}$. Note that the upper limits in Figs.~\ref{Scalar_3} and \ref{Scalar_4} were obtained assuming the NWA. However, when the couplings become large ($g_{i}\gg1$), the resonance can become significantly broad, making the NWA no longer reliable. Moreover, threshold effects can spoil the NWA in the large resonance mass region \cite{Grojean:2011vu}. For these reasons the limit in the high mass region can differ significantly from those in Figs.~\ref{Scalar_3} and \ref{Scalar_4}.

\section{Heavy quark $U$}
\begin{figure}[t]
\centering
\includegraphics[scale=0.24]{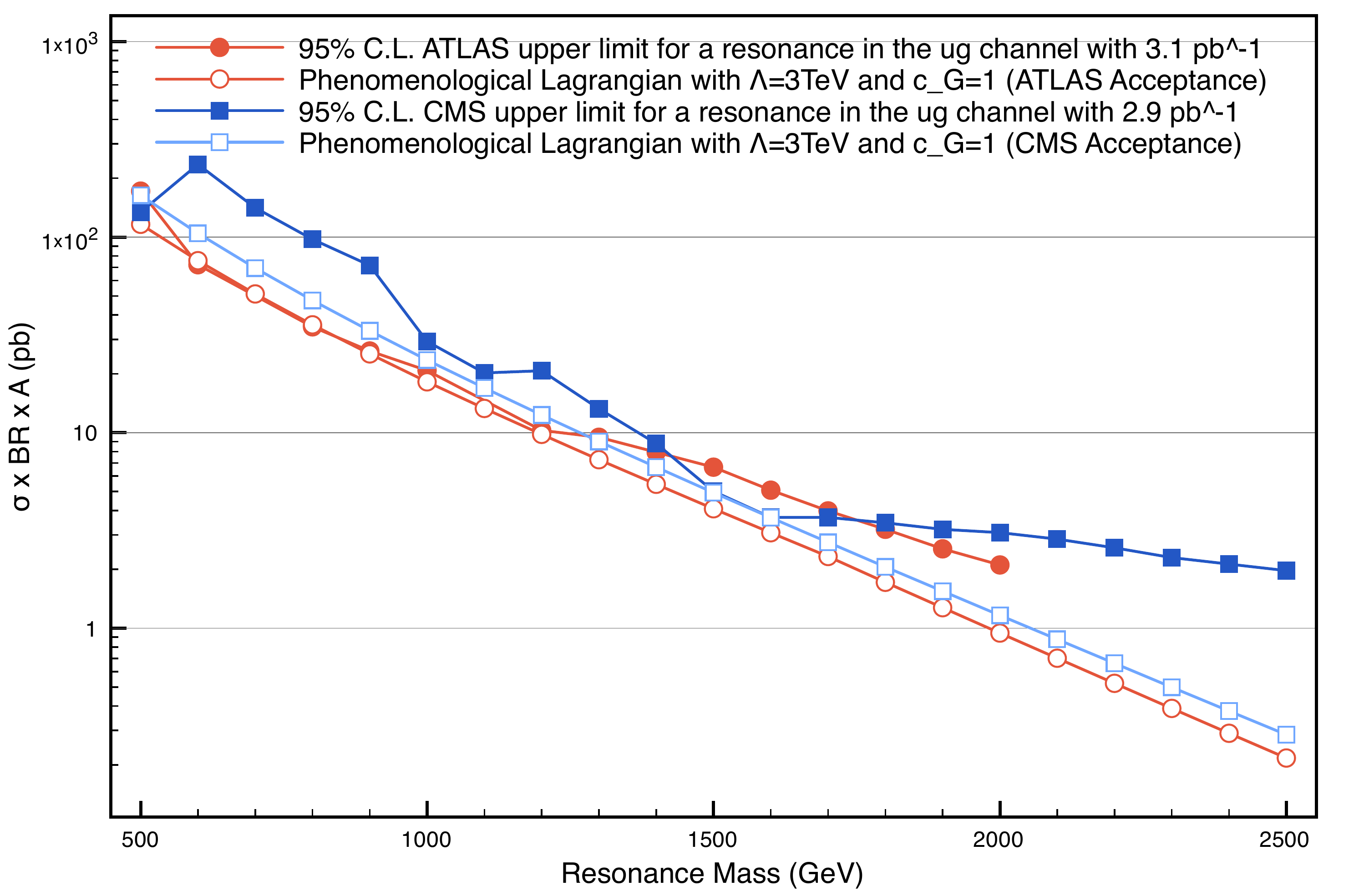}\vspace{3mm}
\includegraphics[scale=0.24]{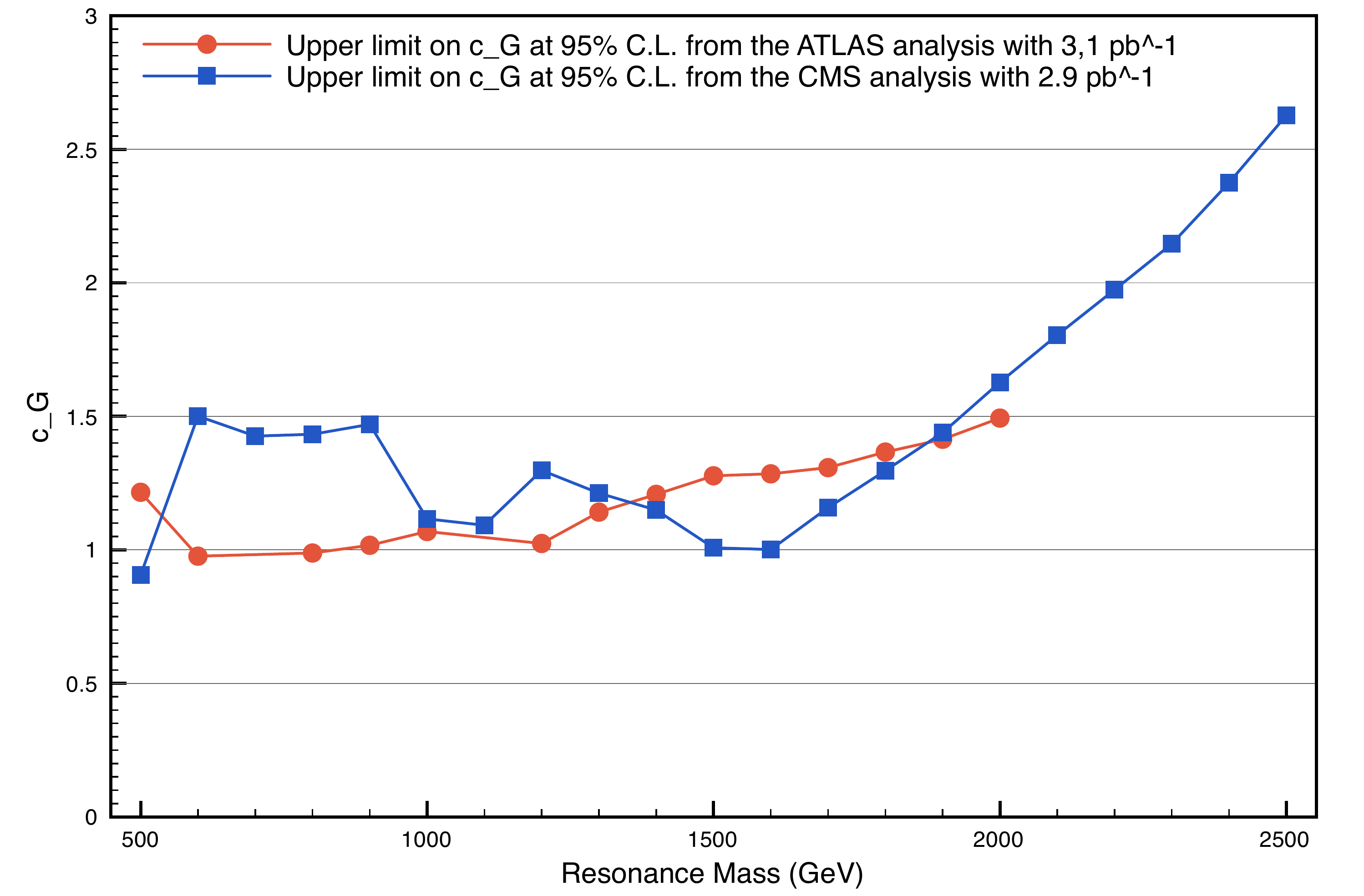}
\caption{Left panel: experimental upper limits at $95\%$ C.L. on the total cross section for a $qg$ resonance measured by the CMS (blue) and ATLAS (red) experiments and cross sections predicted with the Lagrangian \eqref{eq2} for the heavy quark $U$, as functions of the resonance mass; Right panel: upper bound at $95\%$ C.L. on the coupling $c_{G}$ as function of the resonance mass for $\Lambda=3$ TeV, assuming $BR\(S\to qg\)\approx 1$.}
\label{Fermion_3}%
\end{figure}

The same pheomenological study we have made for the scalar singlet can be done for an heavy quark $U$ trasforming as a $\(3,1\)_{2/3}$ of the SM gauge group and coupling to the $qg$ initial state. A reasonalble phenomenological Lagrangian is
\be\l{eq2}
\La_{U}=c_{G}\f{g_{S}}{\Lambda}\bar{U}_{L}\s^{\mu\nu}T^{a}u_{R}G_{\mu\nu}^{a}+c_{B}\f{g^{\prime}}{\Lambda}\bar{U}_{L}\s^{\mu\nu}u_{R}B_{\mu\nu}\,,
\ee
where $\s^{\mu\nu}=i/2 [\g^{\mu},\g^{\nu}]$ and $T^{a}$ are the generators of the fundamental representation of $SU\(3\)$ ($T^{a}=\lambda^{a}/2$). As in the case of $S$, $U$ could be a composite state generated by a strong dynamics responsible for EWSB, in which case NDA suggests $\Lambda\approx 4\pi v\approx 3$ TeV and $c_{G}\approx c_{B}\approx 1/4\pi$. Problems with flavor can arise from the Lagrangian \eqref{eq2} and can be solved by making further assumptions on the new physics \cite{Barbieri:2011p2759}. We don't discuss them here since they don't affect the phenomenology we are interested in.
For the reference values of the parameters $\Lambda=3$ TeV and $c_{G}=c_{B}=1$, the width of the heavy quark $\G_{U}$ ranges from about $2$ GeV for $M_{U}=500$ GeV to about $30$ GeV for $M_{U}=1.5$ TeV.\\
\indent We have made a preliminary study of the sensitivity to the search for $U$ in the di-$jet$ and $\gamma+jet$ final states (see Ref.~\cite{Barbieri:2011p2759} for more details).
We have found that the di-$jet$ channel is again strongly disfavored with respect to the one containing a photon. A discovery in the latter channel looks possible with a modest integrated luminosity of about $5-10$ pb$^{-1}$. Assuming the dominance of the $U\to qg$ channel and the validity of the NWA, an absence of signal for a given integrated luminosity is easily translated in a lower bound on $\Lambda/c_{G}$ for the di-$jet$ channel and $\Lambda/c_{B}$ for the $\gamma+jet$ channel.\\
\indent In this case we have used both the searches of the CMS \cite{CMSCollaboration:2010p2595} and the ATLAS \cite{ATLASCollaboration:2010p1746} experiments in the di-$jet$ channel to set an upper bound on the coupling $c_{G}$ (for fixed $\Lambda=3$ TeV) as a function of $M_{U}$. The upper bounds on the total cross section compared with the prediction for our reference values of the parameters and the bounds on the corresponding coupling $c_{G}$ are depicted in Fig.~\ref{Fermion_3}.%
\begin{figure}[t!]
\centering
\includegraphics[scale=0.26]{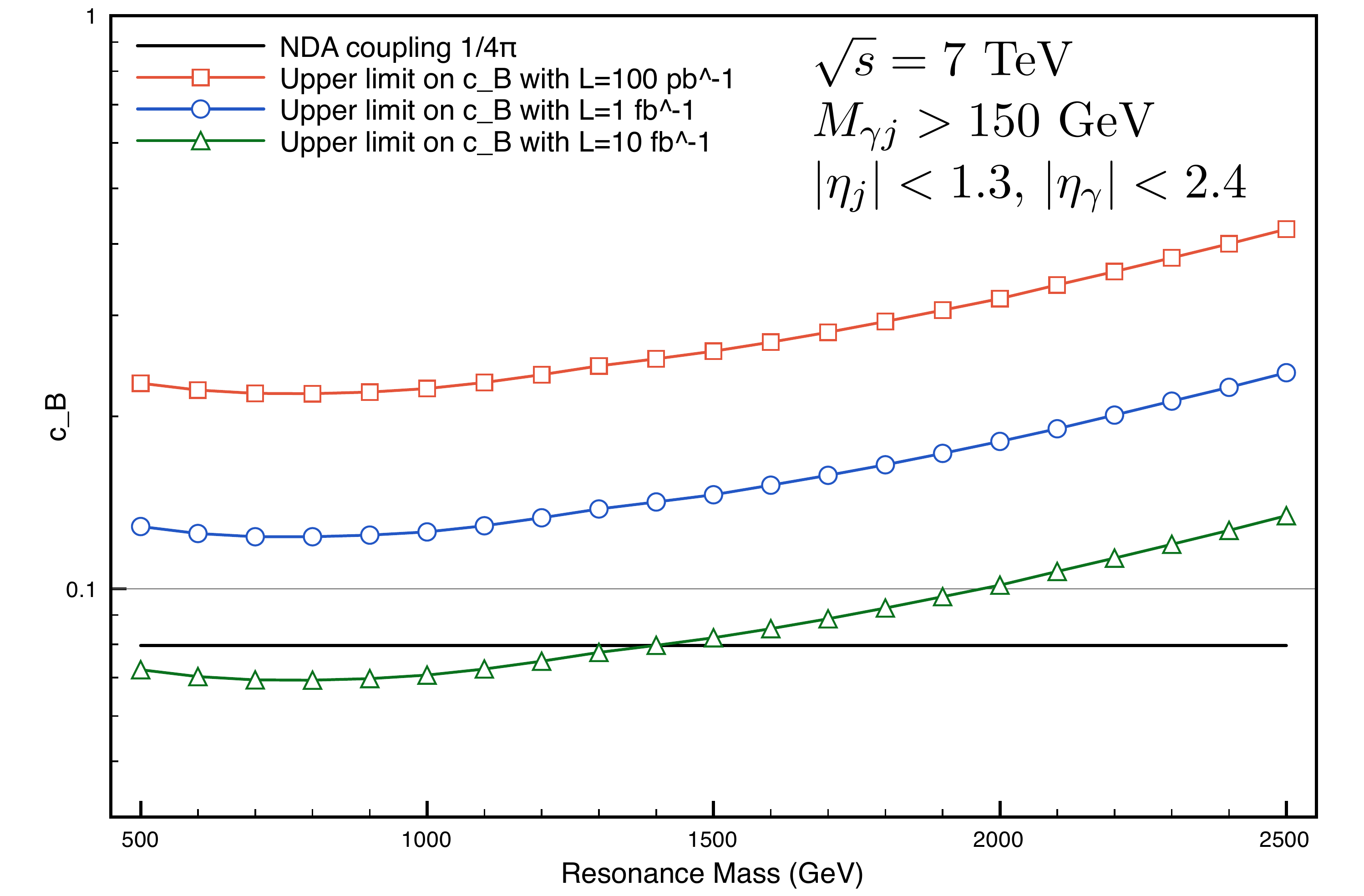}
\caption{Upper limits on $c_{B}$ as functions of the heavy quark mass $M_{U}$ for $\Lambda=3$ TeV assuming $BR\(S\to q\g\)\ll BR\(S\to qg\)$.}
\label{Fermion_4}
\end{figure}%
In Fig.~\ref{Fermion_4} we also show a preliminary estimation of the senstivity of the early LHC to the coupling $c_{B}$ relevant for the $\gamma+jet$ final state. From this figure we can see that the NDA coupling $1/4\pi$ can be excluded, in a wide region of $M_{U}$, with an integrated luminosity of about $1$ to $10$ fb$^{-1}$.

\section{Acknowledgments}
	This work was done in collaboration with Riccardo Barbieri. I want to thank Michelangelo Mangano, Ennio Salvioni and Paolo Francavilla for useful discussions. This work was partially supported by the European Program {\it Unification in the LHC Era}, contract PITN-GA-2009-237920 (UNILHC) and by the European Program {\it Advanced Particle Phenomenology in the LHC Era}, contract PITN-GA-2010-264564 (LHCPhenoNet).

\def\bstname{fdp}

%
 \bibliographystyle{JHEP}
\bibliography{paper}{}
\end{document}